\documentclass[twocolumn]{revtex4}    
\usepackage{amsmath,amssymb,amsthm}
\usepackage{bbold}

\usepackage{graphicx}

\begin{document}

\title{Composite Cluster States and Alternative Architectures for One-
Way Quantum Computation}
\author{Darran F. Milne, and Natalia V. Korolkova}
\email{dfm@st-andrews.ac.uk}
\affiliation{School of Physics and Astronomy, University of St. 
Andrews, North Haugh, St. Andrews, Fife, KY16 9SS, Scotland}

\date{\today}

\begin{abstract}
We propose a new architecture for the measurement-based quantum 
computation model. The new design relies on small composite light-atom 
primary clusters. These are then assembled into cluster arrays using 
ancillary light modes and the actual computation is run on such a 
cellular cluster. We show how to create the primary clusters, which 
are Gaussian cluster states composed of both light and atomic modes. 
These are entangled via QND interactions and beamsplitters and the 
scheme is well described within the continuous-variable covariance 
matrix formalism. 

\end{abstract}
\maketitle

\section{Introduction}

The one-way model or Measurement Based Quantum Computation (MBQC) has 
emerged as a conceptually interesting and potentially practical 
alternative to the standard model of quantum computation 
\cite{Raussendorf}. MBQC replaces the need for coherent unitary 
control \cite{Nielsen} by a sequence of adaptive local measurements 
performed on a highly entangled resource state known as a 
\textit{cluster state} \cite{Briegel}. This resource acts as a 
universal substrate with quantum information encoded virtually within 
it. Though originally based on qubits, the cluster model has been 
generalized to higher dimensional discrete-variable systems (qudits) 
\cite{Zhou} as well as to continuous quantum variables 
\cite{Menicucci}.

In its continuous-variable (CV) incarnation, the resource state is a 
multimode squeezed Gaussian state. In the optical setting 
\cite{vanLoock,Zhang}, homodyne detection and photon counting plus 
classical feedforward, suffice to implement universal QC over CV. It 
has also been shown that homodyne detection alone is sufficient to 
implement all multimode Gaussian operations, given a cluster state 
with a sufficiently connected graph  \cite{Gu}.

Many proposals for the construction of optical CV cluster states have 
been put forward, including the linear optical construction 
\cite{vanLoock}, the single optical parametric oscillator (OPO) method 
\cite{Flammia,Pfister} and single quantum nondemolition (QND)-gate 
schemes \cite{Ralph,Menicucci2}. These procedures are deterministic 
and have the advantage over their discrete counterparts 
\cite{Nielsen2,Browne,Duan}, which rely on nondeterministic 
interactions and postselection. However, as with any CV system, these 
protocols suffer the usual problems such as the lack of infinitely 
squeezed resources leading to finite squeezing errors. 

Here we will use the canonical generation method where
cluster states are created from single-mode squeezers and controlled-Z 
($C_Z$) 
gates \cite{Menicucci}. The controlled-Z is an example of a QND 
interaction and, for optical modes, can be implemented using 
beamsplitters and inline or 
offline squeezers \cite{Braunstein,Yurke,Filip}. This method, while 
experimentally challenging is achievable with current technology 
\cite{Yoshi}.

CV cluster states can be built, not only from optical modes but also 
from ensembles of polarized atoms where each ensemble is a different 
CV mode \cite{Milne1,Milne2,Sanpera1,Sanpera2}. The ensembles are 
entangled by performing interactions with off-resonant linearly 
polarized light, which, on the classical level, performs a Faraday 
rotation. The rotated light then serves as a carrier to encode 
information in each of the ensembles and homodyne measurements are 
performed to complete the protocol and project the state of the 
ensembles into the desired entangled state.

However, it has been shown that Gaussian cluster states suffer from a 
fundamental problem. Namely, no matter what Gaussian local 
measurements are performed on systems distributed on a general graph, 
transport and processing of quantum information is not possible beyond 
a certain influence region, except for exponentially suppressed 
corrections \cite{Eisert}. In other words, as the size of the cluster 
increases there is an exponential decay in the Gaussian localizable 
entanglement over the entire cluster which leads to the corollary that 
cluster states with only Gaussian operations cannot serve as perfect 
quantum wires. Hence large cluster states with only Gaussian resources 
available will always suffer from large errors.

Here we propose an architecture of a one-way computer that removes the 
need for large clusters. In our scheme, small, single gate-size 
cluster states composed of atomic ensembles are created and the 
computations are carried out over an array of such clusters. Due to 
their proven suitability as information carriers, communication 
between the clusters is accomplished using optical modes. The atomic 
ensembles we use have the advantage that they can serve as a short 
term quantum memory as well as a processing device due to their 
relatively long lived storage times. In order to get information from 
one processor to the next, we entangle a polarized light pulse to the 
ensemble. This light mode in effect becomes part of the processor 
cluster state.
When suitable measurements are performed on the 
atomic modes, the information is passed onto the light. This light 
mode can then be coupled to the next processor and the procedure is 
repeated until the computation is complete. In order to realize this 
scheme, we require a protocol to include a light mode into an atomic 
cluster state. We call such a state a \textit{composite cluster state} 
that is composed of $m$ atomic modes and $n$ light modes. Here we will 
explicitly demonstrate the protocol for the generation of such states.

This paper is organized as follow. In section II we will review the 
types of interaction we will be using and show how the two-mode 
composite 
cluster state is constructed. In section III, we discuss the protocol 
in the covariance matrix formalism. In section IV we discuss the 
entanglement properties of the state. 
In section V we generalize our discussion to multipartite entanglement 
and show how to create composite cluster states of arbitrary size. In 
section VI we discuss an alternative computational architecture for CV 
MBQC, making use of the multi-mode composite cluster states. We 
conclude in section VII.
 
\section{Quantum non-demolition interactions}

The physical systems of interest to us are atomic ensembles and light 
beams. The light beams are used both as part of the required cluster 
state and information carriers between nodes of the cluster.

Each atomic ensemble contains a large number, $N_{at}$, of non-
interacting alkali atoms with individual total angular momentum 
$\hat{\textbf{F}}$ \cite{Polzik}. The ensemble is described by its 
collective angular momentum $\hat{\textbf{J}}=(\hat{J}_x, \hat{J}_y, 
\hat{J}_z)$, where 
\begin{equation}
\hat{J}_k = \sum_{i=1}^{N_{at}} \hat{F}_{k,i}, \;\;\; k=x,y,z.
\end{equation}
All atoms are assumed to be polarized along the $x$-direction, which 
corresponds to preparing them in a particular hyperfine state, 
$|F,m_F\rangle$. Then, fluctuations in the $\hat{J}_x$ component of 
collective spin are kept extremely small relative to the strong 
coherent excitation and we can just treat $\hat{J}_{x}$ as a classical 
quantity, $\hat{J}_x \approx \langle \hat{J}_x \rangle \equiv \hbar 
J_x = \hbar N_{atom}F$. The orthogonal components of spin are 
unaffected by the $x$-polarization and quantum fluctuations around 
their mean value remain relatively large. By taking an appropriate 
normalization, the orthogonal components fulfill the canonical 
commutation relations, $[\hat{J}_y / \sqrt{\hbar J_x} , \hat{J}_z / 
\sqrt{\hbar J_x}]=i \hbar $. In this canonical form, we can identify 
these variables as the ``position" and ``momentum" of the system, 
defined in the following way,
\begin{equation}
\hat{x}_A = \frac{\hat{J}_y}{\sqrt{\hbar J_x}},\;\;\; \hat{p}_A = 
\frac{\hat{J}_z}{\sqrt{\hbar J_x}}.
\end{equation}
Light modes are used as a node of the cluster and for generating 
interactions. Each light mode is taken to be out of resonance from any 
relevant atomic transition and linearly polarized along the $x$-
direction. A useful description of light is given through the Stokes 
operators, $\hat{\textbf{s}}=(\hat{s}_x,\hat{s}_y,\hat{s}_z)$ of light 
polarization given by
\begin{align}
\hat{s}_x &= \frac{\hbar}{2}(\hat{n}_x - \hat{n}_y), \nonumber \\
\hat{s}_y &= \frac{\hbar}{2}(\hat{n}_{\nearrow} - \hat{n}_{\searrow}), 
\nonumber \\
\hat{s}_z &= \frac{\hbar}{2}(\hat{n}_{L circ} - \hat{n}_{R circ}).
\end{align}
The individual components correspond to the difference between the 
number of photons (per unit time) with $x$ and $y$ polarization, $\pm 
\pi/4$ linear polarizations and the two circular polarizations, 
respectively. These allow for a microscopic description of the 
interaction with atoms, however only the macroscopic observables 
$\hat{S}_k = \int_{0}^{T} \hat{s}_k(t) dt$, where $T$ is the duration 
of the light pulse will be relevant. Similar to the atomic case, the 
linear polarization along $x$ allows us to make the approximation 
$\hat{S}_x \approx \langle \hat{S}_x \rangle \equiv N_{ph} \hbar /2$. 
The orthogonal components $\hat{S}_y$ and $\hat{S}_z$ are rescaled to 
fulfill the commutation relation $[\hat{S}_y /\sqrt{\hbar 
S_x},\hat{S}_z /\sqrt{\hbar S_x}]=i\hbar$. Once again we make a 
connection with the canonical position and momenta,
\begin{equation}
\hat{x}_L = \frac{\hat{S}_y}{\sqrt{\hbar S_x}}, \;\;\; \hat{p}_L = 
\frac{\hat{S}_z}{\sqrt{\hbar S_x}}.
\end{equation}
Note, in the language of the canonical variables, the spin and 
polarization degrees of freedom are treated on an equal footing. In 
the following analysis, we will only use these canonical variables to 
refer to the atomic ensembles and light modes We denote by $A_i (L_i)$ 
the $i^{th}$ atomic ensemble (light mode).
Let us assume that a light beam propagates in the $YZ$ plane and 
passes through a single ensemble at an angle $\alpha$ with respect to 
the $z$-direction. Then the atom-light interaction can be approximated 
by the effective QND Hamiltonian \cite{Julsgaard},
\begin{equation}
H_{eff}(\alpha) = \frac{\kappa}{T} \hat{p}_L(\hat{p}_A \cos \alpha + 
\hat{x}_A \sin \alpha),
\end{equation}
where $\kappa$ is the coupling constant. The evolution associated with 
this interaction can be evaluated by applying the Heisenberg equation 
for the atoms and the Maxwell-Bloch equation (neglecting retardation) 
for the light. The variables characterizing the composite system 
transform according to the following rules:
\begin{align}
\hat{x}^{out}_A &= \hat{x}^{in}_A - \kappa \hat{p}^{in}_L \cos \alpha, 
\nonumber \\
\hat{p}^{out}_A &= \hat{p}^{in}_A - \kappa \hat{p}^{in}_L \cos \alpha, 
\nonumber \\
\hat{x}^{out}_L &= \hat{x}^{in}_L - \kappa (\hat{p}^{in}_A \cos 
\alpha+ \hat{x}^{in}_A \sin \alpha), \nonumber \\
\hat{p}^{out}_L &= \hat{p}_L.
\end{align}
These are straightforwardly generalized to the case in which a single 
light beam propagates through many atomic ensembles, impinging on the 
$i^{th}$ sample at an angle $\alpha_i$.

We also require an interaction between the cluster light mode and the 
interaction light pulses. These interactions can be implemented by a 
beamsplitter and squeezers \cite{Braunstein}, yielding an interaction 
Hamiltonian of the form $H = \hat{x}_L \hat{x}_i$, where $\hat{x}_i$ 
is the position quadrature of the interaction light pulse. Under the 
influence of this Hamiltonian the variables transform according to:
\begin{align}
\hat{x}^{out}_L &= \hat{x}^{in}_L , 									
\nonumber \\
\hat{p}^{out}_L &= \hat{p}^{in}_L - \hat{x}^{in}_i ,  \nonumber \\
\hat{x}^{out}_i &= \hat{x}^{in}_i   , 								
\nonumber \\
\hat{p}^{out}_i &= \hat{p}_i- \hat{x}^{in}_L.
\end{align}
Note that $\hat{p}_i$ ($\hat{x}_i$) are the squeezed (antisqueezed) 
modes such that, $\hat{p}_{L/i} = e^{-r}\hat{p}^{(0)}_{L/i}$ and 
$\hat{x}_{L/i} = e^{+r}\hat{x}^{(0)}_{L/i}$ with vacuum modes labelled 
by the superscript $(0)$.

\subsection{Composite cluster protocol}

The stabilizer formalism gives us an efficient way to represent 
continuous variable cluster states. A state $|\phi \rangle$ is 
stabilized by an operator $K$ if it is an eigenstate of $K$ with unit 
eigenvalue. That is $K|\phi \rangle = |\phi \rangle$. If such a set of 
operators exist for a given state, we call that state a stabilizer 
state and we may use the generators of its stabilizer group to 
uniquely specify it. It is well known that the stabilizers for 
continuous variable cluster states are 
\begin{equation}
K_{i}(S)=X_{i}(S) \prod_{j\in N(i)} Z_j(s), \;\;\;\; i=1,...,n
\end{equation}
for all $s \in \mathbb{R}$, where $N(i)$ is the set of vertices that 
neighbour vertex $i$, i.e. $N(i)=\{j |(v_i,v_j)\in E\}$. This group is 
described by its Lie algebra, the space of operators $H$ such that 
$H|\phi\rangle=0$. We refer to any element of this algebra as a 
nullifier of $|\phi\rangle$. Being hermitian, every nullifier is an 
observable \cite{Knill}. Any ideal cluster state has nullifier 
representation
\begin{equation}
H_i=\hat{p}_i - \sum_{j \in N(i)}\hat{x}_j, \;\;\;\; i=1,...,n.
\end{equation}
where $H_i \rightarrow 0$ for an ideal cluster. 

For the case of the two mode composite cluster composed from an atomic 
ensemble and optical mode, these nullifiers reduce to just two 
conditions on the quantum variables,
\begin{equation}\label{conditions}
\hat{p}_A - \hat{x}_L \rightarrow 0, \;\;\;\; \hat{p}_L - \hat{x}_A 
\rightarrow 0
\end{equation}
where the subscript $A$ refers to the atomic ensemble and $L$ is the 
light mode. Using the interaction pulses we can assemble these 
quadrature combinations in the following way. Unlike standard two-mode 
entangled states, the two-mode cluster state mixes the position and 
momentum quadratures of different nodes. 

\begin{figure}[htp]
\begin{center}
\includegraphics[width=8.5cm]{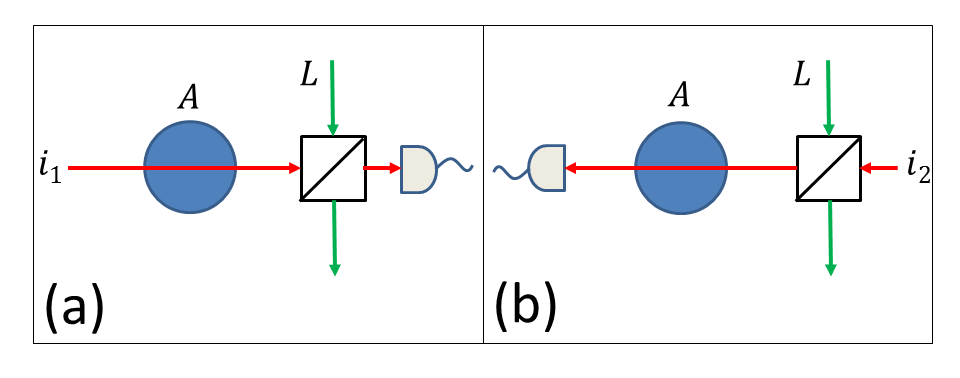}
\end{center}
\caption{Schematic description of the creation of a two mode composite 
cluster state. Light modes can be combined with atomic ensembles to 
form the cluster states by enacting a QND interaction with squeezers 
and beamsplitters. (a) The first interaction encodes atomic 
quadratures onto the light mode $L$. (b) The second interaction 
encodes light quadratures onto the atomic ensemble $A$. These 
interactions are required to fulfil the nullifier condition of Eq. 
(\ref{conditions})/}
\label{protocol1}
\end{figure}

We pass our first interaction pulse, $i_1$ though the atomic ensemble 
(see Fig.\ref{protocol1}(a)). By careful choice of the angle at which 
the light impinges on the atomic sample we can couple the light and 
the atoms via the QND Hamiltonian $H=\kappa \hat{x}_A \hat{p}_{i_1}$. 
The interaction pulse picks up an atomic quadrature term in its 
position variable, $\hat{x}'_{i_{1}}=\hat{x}^{in}_{i_{1}} + \kappa 
\hat{x}^{in}_A$. Then the interaction pulse is combined with the 
cluster light mode, $L$, on a beamsplitter with effective interaction, 
$H=\hat{x}_{L}\hat{x}_{i_{1}}$. This modifies the quadratures of the 
light mode by rotating $\hat{p}_L$ to $\hat{p}'_{L}=[\hat{p}_L^{in}-
\kappa \hat{x}_A^{in}]- \hat{x}^{in}_{i_{1}}$. 
Our second interaction pulse $i_2$ is first passed through the 
beamsplitter with light mode $L$ (Fig.\ref{protocol1}(b)). The 
quadrature action on $i_2$ is a rotation of $\hat{p}_{i_2}$, 
$\hat{p}'_{i_{2}} = \hat{p}^{in}_{i_{2}} - \hat{x}_L$ with the 
associated backaction on the light mode. Interaction of the pulse with 
the atomic ensemble via $H=\kappa \hat{x}_A \hat{p}_{i_2}$ results in 
the output $\hat{p}^{out}_{A} =[\hat{p}^{in}_A - \kappa \hat{x}_L] - 
\kappa (\hat{p}_{i_2} - \hat{x}_{i_1})$.
Note that the momenta of the light and atomic modes now have the 
correct form as given by (\ref{conditions}) for the quadrature 
combinations for a two mode cluster state plus some extra noise terms 
that are a consequence of the backaction of the QND interactions. To 
complete the protocol, the outgoing interaction pulses undergo 
homodyne measurements which project the composite system into a 
cluster state. As we shall see in section VI, such composite states 
provide the means for the atomic clusters to communicate with each 
other so 
they can act as one computational array.

Due to the fact that both light and atoms are highly polarized, the 
initial states of the atomic ensembles as well as the light can be 
treated as Gaussian modes. The interaction is a bilinear coupling 
between Stokes operators and collective spin and thus preserves the 
Gaussian character of the initial states. Furthermore, the 
beamsplitter interactions that we use to include light modes in the 
cluster also yield a description in terms of Gaussian operations. To 
tackle these CV interactions we employ the \textit{covariance matrix} 
(CM) formalism. This allows us to examine the entanglement properties 
of the resulting state and estimate bounds on the experimental 
implementation.

\section{Symplectic Description}

Gaussian functions are mathematically completely defined by their 
first and second moments \cite{Giedke,vanLoock2,Adesso}. Hence it 
follows that any Gaussian state $\rho$ is characterized by the first 
and second moments of the quadrature field operators. We denote the 
first moments by 
$\bar{R}=(\langle \hat{R}_1 \rangle,\langle \hat{R}_2 
\rangle,...,\langle \hat{R}_N \rangle,\langle \hat{R}_n \rangle)$ and 
the second moments by the covariance matrix (CM) $\sigma$ of elements
\begin{equation}
\sigma_{ij}=\frac{1}{2} \langle \hat{R}_i \hat{R}_j+ \hat{R}_f 
\hat{R}_i \rangle - \langle \hat{R}_i \rangle  \langle \hat{R}_j 
\rangle.
\end{equation}
First moments can be arbitrarily adjusted by local unitary operations, 
that is, displacements in phase space. These can be performed by 
applying single-mode Weyl operators to re-center the reduced Gaussian 
corresponding to each single mode. Such operations leave the structure 
of the Gaussian state and hence any information contained within it 
invariant and hence any first moments are unimportant to our analysis 
and from now on, unless otherwise stated we will set these to zero.

The Wigner function of a Gaussian state can be written as follows in 
terms of phase-space quadrature variables
\begin{equation}
W(R)=\frac{e^{-\frac{1}{2} R \sigma^{-1} R^T}}{\pi \sqrt{Det \sigma}}
\end{equation}
where $R$ is the real phase-space vector 
$(\hat{x}_1,\hat{p}_1,...,\hat{x}_N,\hat{p}_N)$. A useful observation 
is that even though the Hilbert space in which the state lives is 
infinite dimensional, a complete description of an arbitrary Gaussian 
state is therefore encoded in the $2N \times 2N$ CM $\sigma$ matrix, 
which we will use to denote the second moments of the Gaussian state 
or the state itself. In the formalism of statistical mechanics, the CM 
elements are the two-point truncated correlation functions between the 
$2N$ canonical continuous variables. 
The covariance matrix corresponding to a quantum state must fulfill 
the 
positivity condition
\begin{equation}
\sigma + i\Omega_N \geq 0
\end{equation}
where
\begin{equation}
\Omega_N = \bigoplus^{N}_{\mu = 1}\Omega, \;\;\; \Omega = 
\left(\begin{array}{cc} 0& 1 \\ -1 & 0 \end{array} \right).
\end{equation}
In general, for QND and beamsplitter interactions, the covariance 
matrix takes the form
\begin{equation}
\sigma = \left(\begin{array}{cc} A& C \\ C^T & B \end{array} \right)
\end{equation}
where the submatrix $A$ corresponds to those modes that are nodes of 
the cluster, $B$ are the light modes responsible for mediating the 
interactions between the nodes within the cluster states and $C$ is 
the correlations between them.

If a Gaussian state undergos a physical unitary evolution that 
preserves its Gaussianity, then the transformation at the level of the 
CM corresponds to a symplectic transformation by a symplectic matrix 
$S$
\begin{equation}
\sigma_{out} = S^T \sigma_{in} S.
\end{equation}
We also require the ability to perform homodyne detections on the 
outgoing interaction pulses. In the CM formalism, assuming an initial 
displacement of zero, the measurement of quadrature $\hat{x}_L$ with 
outcome $z_L$ leaves the system in the state described by the CM
\begin{equation}
A' = A - C(X B X)^{-1}C^T
\end{equation}
with the displacement
\begin{equation}
d_A = C(X B X)^{-1}(z_L,0),
\end{equation}
where the inverse is understood as the Moore-Penrose pseudo-inverse 
whenever the matrix is not full rank and $X$ is a diagonal matrix with 
the same dimension as $B$ with diagonal entries 
$(1,0,1,0,...,1,0)$.

We assume the initial state of the composite system is given by the 
covariance matrix for the atoms, light and interaction pulses, 
$\sigma_{in}= \mathbb{1}^A_2 \oplus  \mathbb{1}^L_2 \oplus 
\mathbb{1}^{i_1}_2 \oplus \mathbb{1}^{i_2}_2$, where the $2 \times 2$ 
identity matrices stand for single modes.

As described in our protocol above, we begin by passing $i_1$ through 
the atomic ensemble and then combine it with the cluster light mode 
via a beamsplitter. The symplectic matrix for this operation is given 
by
\begin{equation}
S_{int_1}=\left(\begin{array} {cccccccc} 1&0&0&0& \kappa &0&0&0 \\ 
0&1&0&0&0&0&0&0 \\ 0& \kappa &1&0&0&-1&0&0 \\ 0&0&0&1&0&0&0&0 \\ 
0&0&0&-1&1&0&0&0 \\ 0&\kappa &0&0&0&1&0&0 \\ 0&0&0&0&0&0&1&0 \\ 
0&0&0&0&0&0&0&1 \end{array} \right).
\end{equation}
The second round of interactions with $i_2$, is given by the 
symplectic matrix
\begin{equation}
S_{int_2}=\left(\begin{array} {cccccccc} 1&0&0&-\kappa &0&0&0&0 \\ 
0&1&0&0&0&0 &1&0\\ 0&0&1&0&0&0&0&-1 \\ 0&0&0&1&0&0&0&0 \\ 
0&0&0&0&1&0&0&0 \\ 0&0&0&-1&0&1&0&0 \\ 0&\kappa &0&0&0&0&1&0 \\ 
0&0&0&0&0&0&0&1 \end{array} \right).
\end{equation}
Then the resulting state is described by the transformation 
$\sigma_{out} = S^T_{int_2}S^T_{int_1} \sigma_{in} 
S_{int_1}S_{int_2}$ (see appendix for explicit form). However, 
entanglement between the light mode and 
atomic ensemble is not produced until the interaction pulses are 
measured. We perform a homodyne measurement on the outgoing pulses 
$i_1$ and $i_2$ in the $x$-basis
with results $z_1$ and $z_2$ to project into the final state described 
by the CM
\begin{equation}
\sigma_{fin}=\sigma^A - \sigma^C(X \sigma^B X)^{-1} \sigma^{C^T}
\end{equation}
where $\sigma^A$ is the upper left $4 \times 4$ matrix corresponding 
to the atomic and light modes, $\sigma^B$ is the lower right $4\time 
4$ matrix representing the interaction pulses and $\sigma^C$ and 
$\sigma^{C^T}$ are the off diagonal matrices with entries 
corresponding to the correlations between the light mode, the atomic 
ensemble and the interaction pulses. The final covariance matrix for 
the two-mode composite state is given in Appendix. A. The final state 
is independent of the measurement outcomes, but they are present in 
the displacement vector. 

\section{Verification of Entanglement}

On completion of the protocol, an analysis if the correlations induced 
between the atomic ensemble and light must be performed. Through the 
CM formalism we have complete access to all the information we need to 
verify entanglement. A structural separability test, which can only be 
applied when the full CM is available is the positive partial 
transpose (PPT) test \cite{Adesso1,Peres, Horodecki}. In phase space, 
any $N$ mode Gaussian state can be transformed by symplectic 
operations in its Williamson diagonal form $\nu$ \cite{Williamson}, 
such that $\sigma = S^T \nu S$, with $\nu=\text{diag}\{\nu_1,\nu_1,...,\nu_n,\nu_N\}$. The set $\{\nu_i\}$ of all positive-
defined eigenvalues of $|i\Omega \sigma|$ constitutes the symplectic 
spectrum of $\sigma$, the elements of which are the symplectic 
eigenvalues which must fulfill the conditions $\nu_i >1$ to ensure the 
positivity of the density matrix associated with $\sigma$. The 
symplectic eigenvalues, $\nu_i$, are determined by $N$ symplectic 
invariants associated with the characteristic polynomial of the matrix 
$|i\Omega \sigma|$. In order to say something about entanglement, we 
recall that the CM's PPT is a necessary and sufficient condition of 
separability for the $(M+N)$-mode bisymmetric Gaussian states with 
respect to the $M|N$ bipartition of the modes. This also holds for 
$(M+N)$-mode Gaussian states with fully degenerate symplectic 
spectrum. Further, for the case of $M=1$, PPT is a necessary and 
sufficient condition for separability of all Gaussian states 
\cite{Simon,Werner}. In phase space, partial transposition corresponds 
to partial time reversal of the CM, or simply, a change of sign of the 
momenta for chosen modes. If $\{\nu_i\}$ is the symplectic spectrum of 
the partially transposed CM $\tilde{\sigma}$, then a $(1+N)$-mode 
Gaussian state with CM $\sigma$ is separable if and only if 
$\tilde{\nu}_i \geq 1$ $\forall i$. If the partially time reversed CM 
does not fulfill the positivity condition, the corresponding state is 
entangled. 
Computing symplectic eigenvalues for the partially transposed CM 
$\tilde{\sigma}_{fin}$, we 
find that the state is indeed entangled since the smallest 
eigenvalue of $|i \Omega \tilde{\sigma}_{fin} |$ is $\tilde{\nu}= 
0.63$ for an 
interaction strength of $\kappa=0.8$. 

\section{Multimode composite cluster states}

Here we extend our analysis to the multipartite case. In general, 
composite clusters can be composed of $m$ atomic modes with $n$ light 
modes. We shall call these $(m,n)$-composite cluster states. 
Here we give the protocol for a $(4,1)$ cluster, which will 
become the basic unit for the computational scheme in the next 
section. 

\begin{figure}[htp]
\begin{center}
\includegraphics[width=4cm]{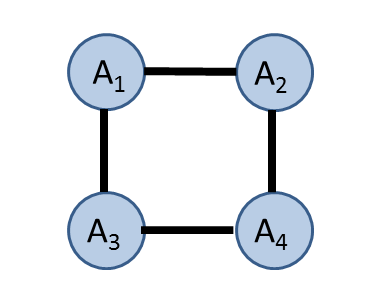}
\end{center}
\caption{A four-node square cluster state, composed of four atomic 
ensembles $A_1,...,A_4$}
\label{square}
\end{figure}

Our protocol for the $(4,1)$-composite cluster proceeds as follows. We 
construct a four-mode square cluster state from atomic ensembles 
labelled $A_i$ (Fig.\ref{square}), where $i=1,...,4$ \cite{Milne1}. We 
include a 
light mode to form the composite state by entangling it with one of 
the atomic modes in the cluster. 

\begin{figure}[htp]
\begin{center}
\includegraphics[width=6cm]{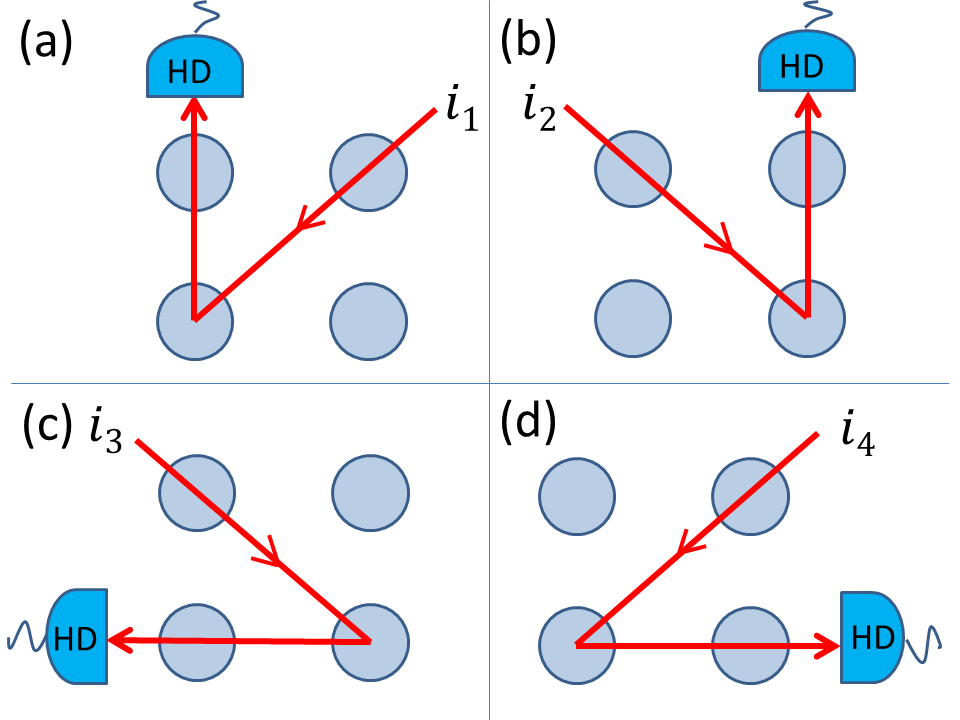}
\end{center}
\caption{Atomic cluster state generation protocol (HD - Homodyne 
detection): (a)Light pulse 
$i_1$ interacts with ensembles $A_2$ and $A_3$, picks up atomic spin 
information which is encoded onto $A_1$. (b)Light pulse $i_2$ 
interacts with ensembles $A_1$ and $A_4$, picks up atomic spin 
information which is encoded onto $A_2$. (c)Light pulse $i_3$ 
interacts with ensembles $A_1$ and $A_4$, picks up atomic spin 
information which is encoded onto $A_3$. (d) Light pulse $i_4$ 
interacts with ensembles $A_2$ and $A_3$, picks up atomic spin 
information which is encoded onto $A_4$. The interaction pulses are 
homodyne measured once these steps have been completed to form the 
cluster state.}
\label{atomprot}
\end{figure}

The nullifiers for a four-mode square cluster can be written as:
\begin{align}\label{conditions}
& \hat{p}_{A_1}-\hat{x}_{A_2}-\hat{x}_{A_3} \rightarrow 0, \;\;\; 
\hat{p}_{A_2}-\hat{x}_{A_1}-\hat{x}_{A_4} \rightarrow 0,\nonumber \\
& \hat{p}_{A_3}-\hat{x}_{A_1}-\hat{x}_{A_4} \rightarrow 0,\;\;\; 
\hat{p}_{A_4}-\hat{x}_{A_2}-\hat{x}_{A_3} \rightarrow 0.
\end{align}
To entangle the ensembles we use light pulses labelled, $i_1,...,i_4$. 
We make 
use of QND Hamiltonians to mediate the interaction between the 
ensembles and light pulses. Note that for such an off-resonant atom-
light interaction, the Hamiltonians are given in \cite{polzik} and are 
well established experimentally. The Hamiltonians we make use of are 
$H_1=\kappa \hat{x}_A\hat{p}_i$ and $H_2=\kappa 
\hat{x}_A \hat{x}_i$ which leads to quadrature transformations
\begin{align}
\hat{x}^{out}_A &= \hat{x}^{in}_A , 									
\nonumber \\
\hat{p}^{out}_A &= \hat{p}^{in}_A - \kappa \hat{x}^{in}_i ,  \nonumber 
\\
\hat{x}^{out}_i &= \hat{x}^{in}_i   , 								
\nonumber \\
\hat{p}^{out}_i &= \hat{p}_i-\kappa \hat{x}^{in}_A.
\end{align}
The protocol to generate atomic cluster states is depicted in 
Fig.\ref{atomprot}. (a) Spin information is picked up from $A_2$ and 
$A_3$ by pulse $i_1$ via the interaction $H_1$. This pulse encodes the 
information onto $A_1$ through the interaction $H_2$. (b) Spin 
information is picked up from $A_1$ and $A_4$ by pulse $i_2$ via the 
interaction $H_1$. This pulse encodes the information onto $A_2$ 
through the interaction $H_2$. (c) Spin information is picked up from 
$A_1$ and $A_4$ by pulse $i_3$ via the interaction $H_1$. This pulse 
encodes the information onto $A_3$ through the interaction $H_2$. 
(d) Spin information is picked up from $A_2$ and $A_3$ by pulse $i_4$ 
via the interaction $H_1$. This pulse encodes the information onto 
$A_4$ through the interaction $H_2$.
When this procedure is completed, the ensembles have quadrature 
combinations:
\begin{align}\label{quad1}
\hat{p}^{out}_{A_1} = \hat{p}_{A_1} - \kappa^2 \hat{x}_{A_2} - 
\kappa^2 \hat{x}_{A_3} - \kappa N_1, \;\;\; 
\hat{x}^{out}_{A_1}=\hat{x}_{A_1}, \nonumber \\
\hat{p}^{out}_{A_2} = \hat{p}_{A_2} - \kappa^2 \hat{x}_{A_4} - 
\kappa^2 \hat{x}_{A_1} - \kappa N_2, \;\;\; 
\hat{x}^{out}_{A_2}=\hat{x}_{A_2}, \nonumber \\
\hat{p}^{out}_{A_3} = \hat{p}_{A_3} - \kappa^2 \hat{x}_{A_4} - 
\kappa^2 \hat{x}_{A_1} - \kappa N_3, \;\;\; 
\hat{x}^{out}_{A_3}=\hat{x}_{A_3}, \nonumber \\
\hat{p}^{out}_{A_4} = \hat{p}_{A_4} - \kappa^2 \hat{x}_{A_2} - 
\kappa^2 \hat{x}_{A_3} - \kappa N_4, \;\;\; 
\hat{x}^{out}_{A_4}=\hat{x}_{A_4}, \nonumber \\
\end{align}
where $N_1 = \hat{x}_{i_1}+\hat{p}_{i_2}+\hat{p}_{i_3}$, $N_2 = 
\hat{x}_{i_2}+\hat{p}_{i_1}+\hat{p}_{i_4}$, $N_3 = 
\hat{x}_{i_3}+\hat{p}_{i_1}+\hat{p}_{i_4}$
and $N_4 = \hat{x}_{i_4}+\hat{p}_{i_2}+\hat{p}_{i_3}$ are the 
backaction terms due to the QND interactions with the interaction 
pulses. Then for $\kappa=1$ the remaining terms are exactly the 
nullifier relations, (\ref{conditions}), for the four-mode square 
cluster state. To complete the protocol, homodyne measurements are 
made on the outgoing interaction pulses which project the ensembles 
into the required state.

\begin{figure}[htp]
\begin{center}
\includegraphics[width=8cm]{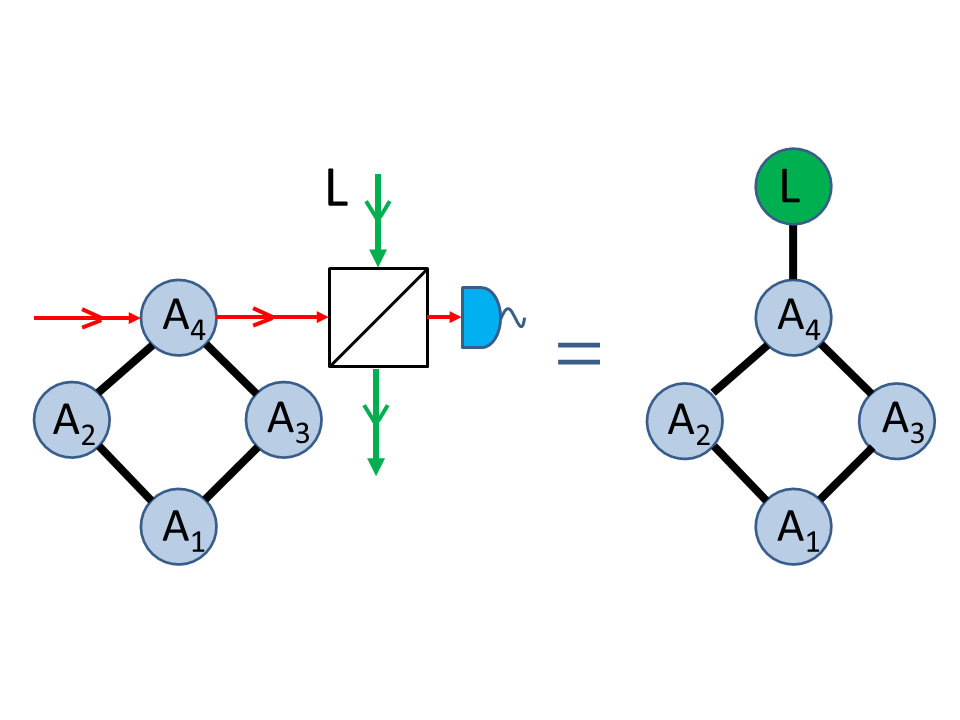}
\end{center}
\caption{To add a light mode to the atomic cluster to form the 
$(4,1)$-composite cluster, we use the protocol of section II. Pulses 
$i_5$ and $i_6$ interact with ensemble $A_4$ and the light mode $L$ 
via a beamsplitter and are subsequently measured.}
\label{comp}
\end{figure}

The $(4,1)$-composite cluster, has nullifier relations
\begin{align}\label{conditions2}
 \hat{p}_{A_1}-\hat{x}_{A_2}-\hat{x}_{A_3}& \rightarrow 0, \;\;\; 
\hat{p}_{A_2}-\hat{x}_{A_1}-\hat{x}_{A_4} \rightarrow 0,\nonumber \\
 \hat{p}_{A_3}-\hat{x}_{A_1}-\hat{x}_{A_4}& \rightarrow 0, \;\;\; 
\hat{p}_{A_4}-\hat{x}_{A_2}-\hat{x}_{A_3} -\hat{x}_{L} \rightarrow 0 
\nonumber \\
 & \hat{p}_L - \hat{x}_{A_4} \rightarrow 0.
\end{align}
We entangle a light mode $L$ with ensemble $A_4$ (Fig.\ref{comp}). 
Following the protocol given in section II, the light mode is 
coupled to the existing atomic cluster with interaction pulses 
$i_5$ and $i_6$. The nullifiers of the atomic modes $A_1$, $A_2$ and 
$A_3$ are unaffected 
but the quadrature combinations for $A_4$ are now
\begin{equation}
\hat{p}^{out}_{A_4}=\hat{p}_{A_4} - \kappa^2 \hat{x}_{A_2} - \kappa^2 
\hat{x}_{A_3} - \kappa^2 \hat{x}_L - \kappa N'_4,
\end{equation}
where $N'_4=\hat{x}_{i_4} +\hat{p}_{i_2}+\hat{p}_{i_3} 
+\hat{p}_{i_5}+\hat{p}_{i_6}$ is the new backaction term. The light 
mode quadratures have transformed as
\begin{equation}
\hat{p}^{out}_L = \hat{p}_L - \kappa^2 \hat{x}_{A_4} - \kappa N_L,
\end{equation}
and $N_L= \hat{x}_{i_5} +\hat{x}_{i_6}$.
Note that the backaction terms, $N_i$, are composed of quadratures of 
the interaction light modes only. The interaction modes are momentum-
squeezed and interact weakly with the ensembles so their backaction 
can be neglected (see also the related experiment \cite{Polzik} where 
this has been verified). We observe that we have a complete set of 
quadrature combinations that satisfies the composite cluster 
nullifier conditions (\ref{conditions2}). Finally, homodyne measuring 
$i_5$ and $i_6$ completes the protocol.

We note that this protocol can be simply extended to creating general 
$(m,n)$-composite clusters of arbitrary shape. Further QND 
interactions can be used to add atomic ensembles and it is always 
possible to add a light mode to an atomic mode through the 
beamsplitter interaction. However adding extra nodes always results in 
extra backaction terms, $N_i$. 

\section{A new architecture}

\begin{figure}[htp]
\begin{center}
\includegraphics[width=8cm]{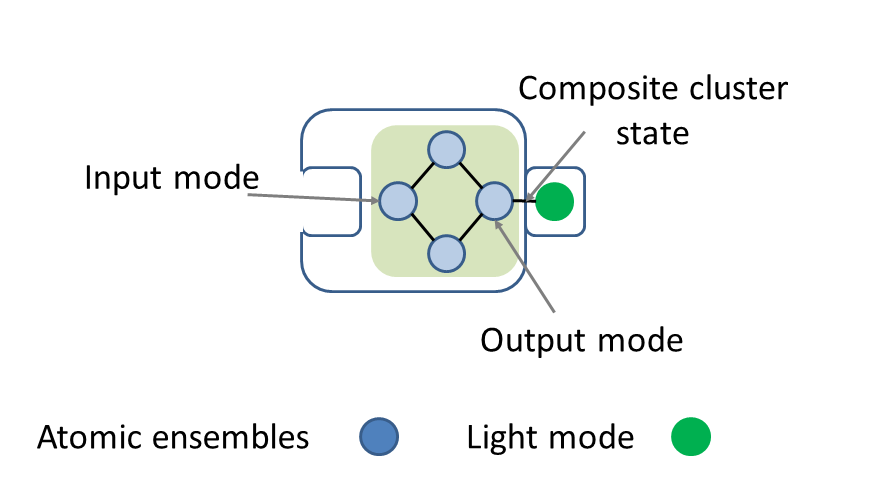}
\end{center}
\caption{A basic \textit{qubrick} composed of a four-mode atomic 
cluster state plus a light mode to create a composite cluster.}
\label{qubrick}
\end{figure}

Here we seek to address one of the major difficulties in the practical 
implementation of cluster state computation: That of decreasing 
localizable entanglement with increasing size of the cluster when only 
Gaussian operations are available \cite{Eisert}. This places a limit 
on the size of useful cluster states, i.e. those that have sufficient 
entanglement available to perform processes below the computational 
error threshold. Traditionally, the MBQC model relies on creating 
large clusters on which the entire computation can be performed and 
read out. Here we propose to build up computational arrays from small
building blocks or \textit{qubricks} which are composed of an atomic 
cluster state acting as a \textit{quantum processor} and a light mode 
that allow for communication to other qubricks (Fig.\ref{qubrick}). 
This allows us to eliminate redundant nodes that only serve to increase 
the error rate in the system. A similar suggestion to increase the 
efficiency of generating qubit cluster 
states based on ancilla light modes was made in \cite{Kendon}.
In our scheme, many atomic cluster states, labelled $(C_1,...,C_N)$, 
are created. Each of these act as a small quantum processor 
which performs some unitary operation $\hat{U}_i$ on an input state. 
Each processor $C_i$, \textit{shares no entanglement} with any other 
in the 
array. We keep the number of nodes in each atomic cluster small, 
say four nodes each (Fig.\ref{square}), which is minimum sufficient 
for a controlled quantum gate. By limiting the size of the cluster 
states, the decay of entanglement within each processor is kept to a 
minimum. 
\begin{figure}[htp]
\begin{center}
\includegraphics[width=8cm]{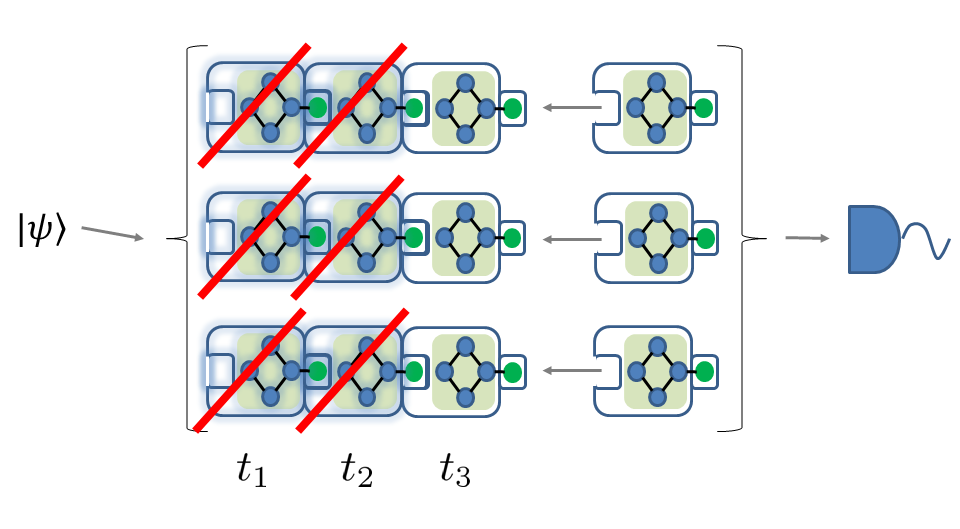}
\end{center}
\caption{The qubricks interact by matching up their output light modes 
with the input of another qubrick. A state can be loaded onto the 
first brick at time $t_1$, the state is processed through adaptive 
measurements and 
the output encodes onto the light mode. The qubrick at $t_1$ having 
been measured no longer contributes to the state. The light pulse is 
connected with the next in the sequence of qubrick processes $t_2$ and 
the process is repeated. This operation can be run in parallel with 
other sequences 
of qubricks and the final state is given by their combined outputs.}
\label{brickarray}
\end{figure}
To allow communication between the processors, we entangle an 
ancillary light mode $L$ with the output node in the atomic cluster 
to form a qubrick which is a $(4,1)$-composite cluster. The light mode 
of the qubrick can then be used to convey the output of the process 
performed on the atomic cluster contained in the  brick to the next in 
the array (Fig.\ref{brickarray}). Given sufficient resources, the 
state can be processed in parallel using strings of qubricks. The 
total output from each of the strings can then be combined to give the 
final output.
\begin{figure}[htp]
\begin{center}
\includegraphics[width=8cm]{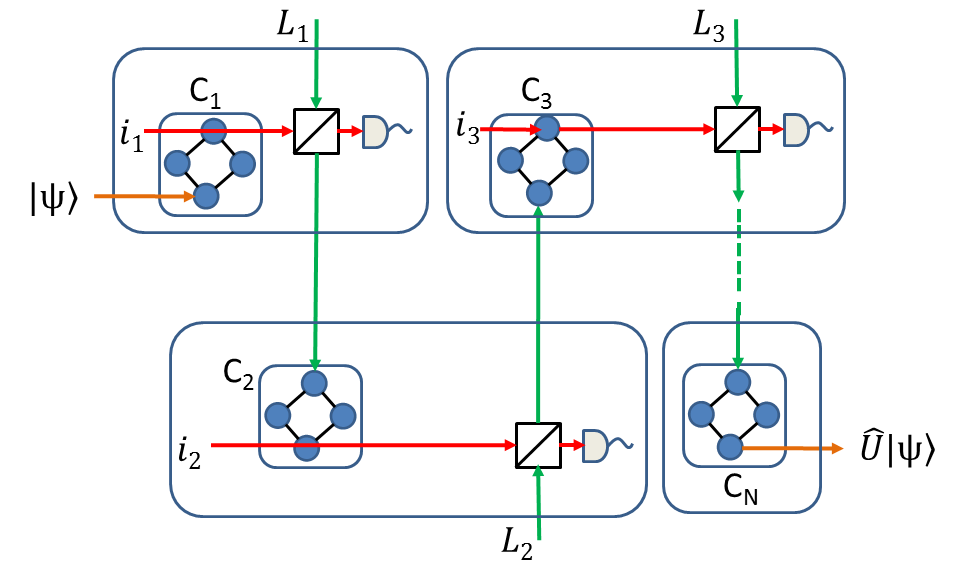}
\end{center}
\caption{Each cluster $C_i$ is kept separate and only communicate 
through light pulses $L_i$. The state to be computed is loaded onto 
$C_1$, which undergos a sequence of adaptive measurements. A composite 
cluster is formed to transfer the state to a light mode which is 
carried to the next cluster in the process. Each cluster acts on the 
state by some unitary $\hat{U}_i$ and the final state is given by 
$\hat{U}|\psi\rangle$.}
\label{arch}
\end{figure}
In more detail, a typical process would proceed as follows. A state 
$|\psi_{ini}\rangle$, is loaded onto the first cluster $C_1$. Adaptive 
measurements are then applied to the atomic ensembles to process the 
state with outcome $|\psi'\rangle = \hat{U}_1|\psi_{ini}\rangle$. A 
light mode $L_1$, is added to $C_1$ to form the qubrick and the state 
is transferred to $L_1$ by a further measurement (Fig.\ref{arch}). 
Since the cluster and its associated entanglement has been destroyed 
in the measurement process the light pulse is free to carry the 
information to the next qubrick, without inadvertently entangling the 
atomic clusters belonging to different bricks. Another measurement 
transfers the state onto the first node of the atomic cluster, $C_2$ 
and a gate operation is performed by a new sequence of measurements 
yielding $\hat{U}_2|\psi'\rangle$. A light mode $L_2$, is added to 
$C_2$ to form a new qubrick and it carries the processed state is to 
the next cluster. Since each cluster applies some unitary 
transformation $\hat{U}_i$ to the state, this process can be repeated 
until the desired output if achieved and the state is given by 
$\hat{U}|\psi\rangle= \hat{U}_i\otimes \hat{U}_{i-1} 
\otimes...\otimes\hat{U}_2\otimes \hat{U}_1|\psi\rangle$. 

It is important to note that we do not claim that this procedure 
outperforms the error rate of large cluster arrays in general. What 
the qubrick scheme does guarantee, is that the loss of entanglement 
and therefore the errors accumulated in each quantum gate are 
constant. Then the errors that propagate through the computation are 
just proportional to the number of qubricks used. This is in contrast 
to the large cluster state in which the entanglement decays 
exponentially with the size of the cluster and hence errors accumulate 
rather quickly as the cluster size increases. Note however, that the 
error rate of particular geometries of large scale cluster states actually 
outperform the qubrick scheme i.e., for a 16-node square cluster. In 
this case the entanglement scales as $C e^{-6}$, where $C$ is some 
constant. If we assume the error rate increases proportional to 
the degradation of entanglement then the errors scale as $a e^{6}$ where $a$ is 
some constant that depends on the particular system in question. In 
the qubrick setting, to replicate this state exactly, we require four 
qubricks and the cumulative error then scales as $a e^{8}$, which is 
significantly worse. Here we do not propose that the qubricks mimic 
the large scale states exactly (since they can never beat the 
fundamental limit anyway). In this scheme, each qubrick and hence each 
quantum gate is initialized individually and only when it is required 
(Fig.(\ref{brickarray})). These gates are then directly coupled to the 
light pulse carrying the output from previous processes. Our 
scheme removes the need for nodes serving as quantum wires to 
convey information around the cluster and eliminates any redundant nodes. 
In doing so we supress the exponential losses due to entanglement 
decay and hence reduce the error rate.

Furthermore, this method can be simplified to just one qubrick, 
if we loop the output from the brick back to the input in a similar manner to that 
proposed in \cite{Ralph}. In this scheme, the output from the first 
computation again gives $|\psi'\rangle =\hat{U}_1|\psi\rangle$. The 
atomic cluster is re-generated and the light mode is fed back into the 
input node where a different series of adaptive measurements is 
performed. This gives $\hat{U}_2|\psi'\rangle$ which is processed 
again until the desired unitary is enacted. This scheme eliminates the 
parallel element but it vastly decreases the number of resources 
required while maintaining a constant error rate. This time the error 
depends on the number of time the state is reused. This type of scheme 
could from the basis of a proof of principle demonstration of CV 
cluster state computation with atomic ensembles since it is achievable 
with currently available technology.

\section{Conclusion}   

We have presented a scheme to produce a two-mode cluster state 
composed of an atomic ensemble and light pulse, called a composite 
cluster, via QND interactions and beamsplitters. We find that since 
both the atomic ensembles and light are initially Gaussian states and 
only Gaussian operations are performed, the procedure is well 
described by the covariance matrix formalism. From the covariance 
matrix we have confirmed that the states are indeed entangled using
the PPT separability criterion. We have generalized our protocol, and 
given an 
explicit construction for multipartite composite cluster states 
composed of an arbitrary number of atomic and light modes. Using our 
composite clusters, we have proposed a new architecture for the one-
way 
computer that reduces losses due to decay of localizable entanglement 
by preparing many separate clusters and using the light modes as 
information carriers between them. This scheme is experimentally 
feasible with current technology and has the potential for scalability 
since many of the resources can be re-used as many times as required.  

\begin{acknowledgements}
This research has been supported by the EU STREP project COMPAS FP7-
ICT-2007-C-212008 under the FET-Open Programme,  
by the Scottish Universities Physics Alliance (SUPA) and by the 
Engineering and Physical Sciences Research Council (EPSRC). 
 \end{acknowledgements}

\appendix 

\section{Explicit covariance matrices for two-mode composite cluster 
state}

Here we give the explicit CM for the two-mode composite cluster state. 
Computing $\sigma_{out} = S^T_{int_2}S^T_{int_1} \sigma_{in} 
S_{int_1}S_{int_2}$ yields the CM:

\begin{widetext}
\begin{equation}
\sigma_{out} = \left( \begin{array}{cccccccc} 3\kappa^2 +1&0&0&-\kappa 
& 2\kappa& 0 &2\kappa &0 \\
																							
0&1&\kappa &0&0&-\kappa & 0& -\kappa \\
																							
0&\kappa & \kappa^2 +3 &0&0& -(\kappa^2+1)&0& -(\kappa^2+1) \\
																							
-\kappa & 0&0& 1&-1&0&-1&0 \\
																							
2\kappa & 0&0&-1&2&0&1&0 \\
																							
0&-\kappa & -(\kappa^2+1) &0&0&\kappa^2+1 &0& \kappa^2 \\
																							
2\kappa & 0&0&-1&1&0&2&0 \\
																							
0&-\kappa & -(\kappa^2+1)&0&0&\kappa^2 & 0&\kappa^2+1 
\end{array}\right)
\end{equation}
\end{widetext}
However, this has not produced the required entanglement between the 
atomic and light mode. Homodyning the interaction pulses produces the 
required correlations. In terms of covariance matrices, this amounts 
to 
the operation
\begin{equation}
\sigma^{A'} = \sigma^A - \sigma^C(X \sigma^B X)^{-1}\sigma^{C^T}
\end{equation}
which traces out the interaction light modes $i_1$ and $i_2$ to leave 
us with a $4\times 4$ matrix representing only the atomic and light 
modes that are part of the cluster:
\begin{widetext}
\begin{equation}
\sigma_{fin} = \left(\begin{array}{cccc} 1+\frac{\kappa^2}{3}& -
\frac{2}{3}\kappa(1+\kappa^2)& -\frac{8\kappa^2}{3}&-\kappa -\frac{2}
{3}\kappa(1+\kappa^2)\\
																				 
-\frac{2}{3}\kappa(1+\kappa^2)& 1+\frac{2}
{3}\kappa(1+\kappa^2)^2&\kappa -\frac{2}{3}\kappa(1+\kappa^2)& 					
\frac{2}{3}\kappa(1+\kappa^2)^2 \\
																				 
-\frac{8\kappa^2}{3}&\kappa -\frac{2}{3}\kappa(1+\kappa^2)&3-
\frac{5\kappa^2}{3}& -\frac{2}{3}\kappa(1+\kappa^2) \\
																				 
-\kappa -\frac{2}{3}\kappa(1+\kappa^2)& \frac{2}
{3}\kappa(1+\kappa^2)^2&-\frac{2}{3}\kappa(1+\kappa^2)&1+\frac{2}{3}
(1+\kappa^2)^2 \end{array} \right)  
\end{equation}
\end{widetext}
This final CM represents the state (up to a known displacement that 
depends on the measurement outcome) of the two-mode composite cluster. 
Here, the upper left $2 \times 2$ block represents the atomic mode, 
the 
lower right $2 \times 2$ is the light mode and the off diagonal blocks 
are the correlations between them.

\end{document}